\documentclass[unnumsec,webpdf,contemporary,large]{oup-authoring-template}

\usepackage{float}
\usepackage{placeins}
\usepackage{xurl}
\usepackage{hyperref}
\begin{document}

\journaltitle{Genomics, Proteomics \& Bioinformatics}
\DOI{DOI added during production}
\copyrightyear{2026}
\pubyear{2026}
\vol{XX}
\issue{x}
\access{Published: Date added during production}
\appnotes{Research Article}

\firstpage{1}

\title[Benchmarking G2P workflows in openSNP]{Benchmarking end-to-end genotype-to-phenotype prediction workflows across 80 openSNP phenotypes}

\author[1,2]{Muhammad Muneeb}
\author[1,2,$\ast$]{David B. Ascher}
\author[1,2]{YooChan Myung}
\author[3]{Samuel F. Feng}
\author[4]{Andreas Henschel}

\address[1]{\orgdiv{School of Chemistry and Molecular Biology}, \orgname{The University of Queensland}, \orgaddress{\street{Queen Street}, \postcode{4067}, \state{Brisbane, QLD}, \country{Australia}}}
\address[2]{\orgdiv{Computational Biology and Clinical Informatics}, \orgname{Baker Heart and Diabetes Institute}, \orgaddress{\street{Commercial Road}, \postcode{3004}, \state{Melbourne, VIC}, \country{Australia}}}
\address[3]{\orgdiv{Department of Science and Engineering}, \orgname{Sorbonne University Abu Dhabi}, \orgaddress{\street{Hazza Bin Zayed Street}, \postcode{38044}, \state{Abu Dhabi}, \country{United Arab Emirates}}}
\address[4]{\orgdiv{Department of Electrical Engineering and Computer Science}, \orgname{Khalifa University of Science and Technology}, \orgaddress{\street{Saada Street}, \postcode{127788}, \state{Abu Dhabi}, \country{United Arab Emirates}}}

\corresp[$\ast$]{Corresponding author. \href{mailto:d.ascher@uq.edu.au}{d.ascher@uq.edu.au}}


\abstract{
Genotype-to-phenotype prediction is a central goal of statistical genetics, yet practical comparisons of prediction workflows remain limited in small, heterogeneous, participant-shared genomic datasets. Here, we benchmarked end-to-end case-control prediction across 80 curated binary phenotypes from openSNP using machine learning, deep learning, and polygenic score workflows. We evaluated 29 machine-learning algorithms, 80 deep-learning model variants, and 3 polygenic score tools across 675 clumping and pruning configurations. No workflow family dominated universally. Polygenic score workflows achieved the highest observed discrimination for 53 phenotypes, whereas machine-learning or deep-learning workflows achieved the highest for 27. However, many apparent phenotype-level wins were modest, with 41.2\% of comparisons representing practical ties within five discrimination points. Performance was strongly phenotype-dependent and sensitive to modeling and preprocessing choices. Distinct workflow-specific failure modes were also observed, including unstable behaviour in PRSice and non-informative collapse in lassosum for 13 phenotypes. Higher peak performance was concentrated in smaller phenotypes, reinforcing the need for cautious interpretation in limited-data settings. The cohort was predominantly of European ancestry, restricting generalisability. Together, these results position openSNP as a useful stress-test environment for genomic prediction and support benchmark-guided workflow selection under realistic conditions of data scarcity, phenotype heterogeneity, and ancestry imbalance.
}


\boxedtext{Key Messages}{
\begin{itemize}
\item We present a large-scale end-to-end benchmark of genotype-to-phenotype prediction workflows across 80 cleaned binary phenotypes from openSNP.
\item No workflow family dominated universally: polygenic score workflows more often achieved the highest observed AUC, but many phenotype-level differences were practically modest.
\item Stability diagnostics, margin-based comparisons, ancestry composition, and workflow-specific failure modes materially improve interpretation beyond simple winner counts.
\end{itemize}}
 
\maketitle
\begin{flushleft}
Total word counts (from ``Introduction'' to ``Materials and Methods''): 5554\\
Total references: 91\\
Total figures: 4\\
Total tables: 10\\
Total supplementary figures: 0\\
Total supplementary tables: 11\\
Total supplementary files: 1\\
Letter counts in title: 81\\
Letter counts in running title: 38\\
Counts in keywords: 6\\
Word counts in abstract: 195
\end{flushleft}

\section{Introduction}

Predicting phenotype from genotype remains one of the central challenges in human genomics, with applications spanning disease-risk stratification, case-control classification, and biological hypothesis generation \cite{Virolainen2022,Dick2011}. The ability to determine phenotypes, especially disease risk, using genomic data is increasingly important in the era of precision medicine \cite{Daneshjou2017,Dong2022}, and the identification of single-nucleotide polymorphisms (SNPs) responsible for variation in phenotypes can help drive new diagnostic and therapeutic options \cite{Ku2010,Keita2004}. Although genome-wide association studies (GWAS) have transformed locus discovery \cite{Uffelmann2021,Shaffer2012}, translating dispersed genetic signal into reliable individual-level prediction remains difficult because most common traits are polygenic, effect sizes are small, and predictive performance depends heavily on phenotype definition, ancestry, sample size, linkage disequilibrium structure, and modelling strategy \cite{Tam2019}. In practice, the key question is often not whether genetic signal exists, but which analytical workflow is most robust for extracting it under realistic data constraints.

A wide range of computational strategies are now used for genotype-to-phenotype prediction. Polygenic risk scores (PRS) remain among the most widely adopted approaches when suitably powered discovery data are available \cite{Choi2020,Lewis2020}, and have been applied to a range of diseases and phenotypes \cite{Collister2022}, including Type-2 diabetes \cite{Merino2022}, attention deficit hyperactivity disorder \cite{DuRietz2018}, and cardiovascular disease \cite{OSullivan2022}. Computational modelling approaches can additionally integrate genomics, transcriptomics, proteomics, and environmental factors to better capture complex interactions \cite{Raimondi2022,Tonner2022,Pires2016}. Machine-learning and deep-learning workflows are attractive because they can capture non-linear relationships, interaction structure, and high-dimensional feature spaces that may not be well represented by additive models alone \cite{Danilevicz2022,Guo2023,Muneeb2022images,Fadista2016,Kaler2019}, and can handle large datasets integrating several genomic data types \cite{Kick2023}. Such pipelines have been applied to humans including facial features \cite{Gurovich2019,doi:10.2147/CCID.S339547}, diseases \cite{BracherSmith2020}, and behavioural phenotypes \cite{McCormack2020}, as well as to plants \cite{Grinberg2019,Ma2018,Shook2021,Singh2016,Gill2022} and animals \cite{Katsaouni2021,Enoma2022,okser2014regularized,taghavi2018deep}. Researchers have employed multiple ML algorithms \cite{Leung2013}, L1L2 regularisation \cite{Guzzetta2010}, deep convolutional neural networks \cite{Liu2019}, multilayer perceptrons \cite{AbdollahiArpanahi2020}, and combinations of random forest and SVM \cite{Gaudillo2019}, alongside genomic selection \cite{Budhlakoti2022}, transcriptomics and proteomics \cite{Wang2019}, and phenome-wide association studies \cite{Bastarache2022}. However, comparative performance across these workflow families is highly context-dependent, and recent studies have emphasised that predictive behaviour is shaped as much by data representation, preprocessing, and validation design as by the learning algorithm itself.

This problem is especially acute in participant-shared and direct-to-consumer genomic datasets. openSNP provides an unusual but valuable resource in which users publicly share genotype data and self-reported phenotypes across a broad range of traits \cite{Greshake2014}. At the same time, such data are analytically challenging: phenotype labels are inconsistent, sample sizes vary widely, genotyping platforms differ, genotype completeness is uneven, covariates are sparse, and ancestry structure is likely to be entangled with platform and missingness artefacts. In a mixed direct-to-consumer dataset combining different genotyping providers, chip designs, missingness patterns, and self-reported phenotypes, a model can learn ancestry, platform, or batch artefacts rather than biology. These features make openSNP poorly suited for claims of clinical deployment, but highly informative as a stress-test environment for genomic prediction workflows under realistic conditions of heterogeneity and limited data. Our study primarily focuses on case-control classification rather than the identification of specific SNPs \cite{Monk2021}, and the comparison of ML and PRS tools is of particular interest since for some phenotypes such as Parkinson's Disease \cite{Cope2021} and breast cancer \cite{Badr2020} ML and DL approaches have produced competitive results, while PRS has performed more strongly for phenotypes such as coronary artery disease \cite{Gola2020}. Deep learning approaches have also been combined with PRS to explore classification performance \cite{Amadeus2021,Zhou2023,Sehrawat2023}.

Several previous studies have explored genotype-phenotype prediction in openSNP, but they have typically focused on a small number of traits or a limited set of modelling approaches. Naret et al.~\cite{Naret2020} calculated PRS for height using ML techniques and achieved an explained variance of 0.53. Saha et al.~\cite{Saha2021} used graph theory to find multi-loci associated with astigmatism. Rajesh et al.~\cite{Rajesh2020} used statistical approaches and ML to predict diabetic macular oedema and achieved an F1 score of 0.84. Lu et al.~\cite{8126853} used a Bayesian network for SNPs and trait inferences for multiple traits. Javier et al.~\cite{733675} proposed a tool for case-control classification using eight ML models. Muneeb et al.~\cite{Muneeb2021} used ML and DL models for eye colour and type-2 diabetes prediction. In all prior studies, researchers considered at most ten phenotypes from openSNP. Ma et al.~\cite{Ma2021} concluded that different PRS methods exhibit distinct performance characteristics across traits, and that performance of the leading model depends on the p-value threshold, clumping, and pruning parameters \cite{Priv2019}. Transfer learning has also been employed in genetics to use knowledge from well-studied populations for prediction in under-studied populations with limited data \cite{Li2022,transferlearning}, and the present results may assist in understanding which workflows are viable when genotype data are limited. Here, we expand the scope substantially by benchmarking end-to-end ML, DL, and PRS workflows across 80 cleaned binary phenotypes \cite{Muneeb2022variation}, asking a more practical question: under phenotype-heterogeneous, small-sample conditions, which workflow families are most often competitive, how large are their practical differences, and what kinds of instability or failure modes emerge?

Here, we benchmarked 29 machine-learning algorithms, 80 deep-learning model variants, and 3 PRS tools with 675 clumping and pruning configurations using openSNP genotype data processed into 80 binary phenotypes \cite{Muneeb2022variation}. For ML and DL workflows, we derived phenotype-specific SNP subsets from training-fold association statistics using p-value thresholding and evaluated predictive performance using repeated stratified train-test splits. For PRS workflows, because external GWAS summary statistics were available for only a subset of phenotypes and often had ancestry mismatch and genotype-rate discrepancies with the openSNP target data, we generated training-derived GWAS summary statistics internally, harmonised them with target genotype data, and compared multiple score-construction strategies. This design constitutes an internal, small-sample, genotype-only score construction benchmark rather than a classical PRS evaluation using large external discovery cohorts, and the manuscript should therefore be interpreted as an exploratory benchmark of end-to-end prediction workflows rather than as a definitive ranking of algorithms in isolation. Across phenotypes, PRS workflows achieved the highest observed AUC for 53 phenotypes while ML/DL workflows achieved the highest observed AUC for 27 phenotypes. These results indicate that no single workflow family dominates universally and that genotype-to-phenotype prediction in small, heterogeneous datasets is highly sensitive to phenotype architecture, feature selection, and workflow design.

\section{Results}
\label{sec:results}

\subsection*{openSNP is a heterogeneous, predominantly European-ancestry stress-test cohort}
\label{sec:popstruct_results}

Principal component analysis of the 1,188 unique openSNP individuals projected onto the 1000 Genomes Phase 3 reference panel revealed that the cohort is predominantly of European ancestry, consistent with the self-reported geographic distribution of openSNP respondents (60.33\% from the USA, 5.17\% from Canada, 4.61\% from the UK). Of the 1,188 individuals, 1,104 (92.9\%) clustered nearest to the European (EUR) superpopulation centroid, while 26 (2.2\%) clustered nearest to East Asian (EAS), 26 (2.2\%) to African (AFR), 17 (1.4\%) to South Asian (SAS), and 15 (1.3\%) to Admixed American (AMR) reference populations. This strong European ancestry enrichment is consistent with the direct-to-consumer genetic testing market, which has historically been concentrated in North American and Western European populations. This ancestry composition represents a key limitation of the benchmark, as the reported performance metrics reflect genotype-to-phenotype prediction primarily within a European-ancestry context, and the findings may not generalise to more diverse populations. Formal within-ancestry sensitivity analyses were not performed due to insufficient sample sizes in non-European ancestry groups.

\subsection*{Benchmarking design compares ML, DL and PRS workflows across 80 phenotypes}
\label{sec:benchmarkdesign_results}

\begin{figure*}[!ht]
\centering
\includegraphics[width=\textwidth]{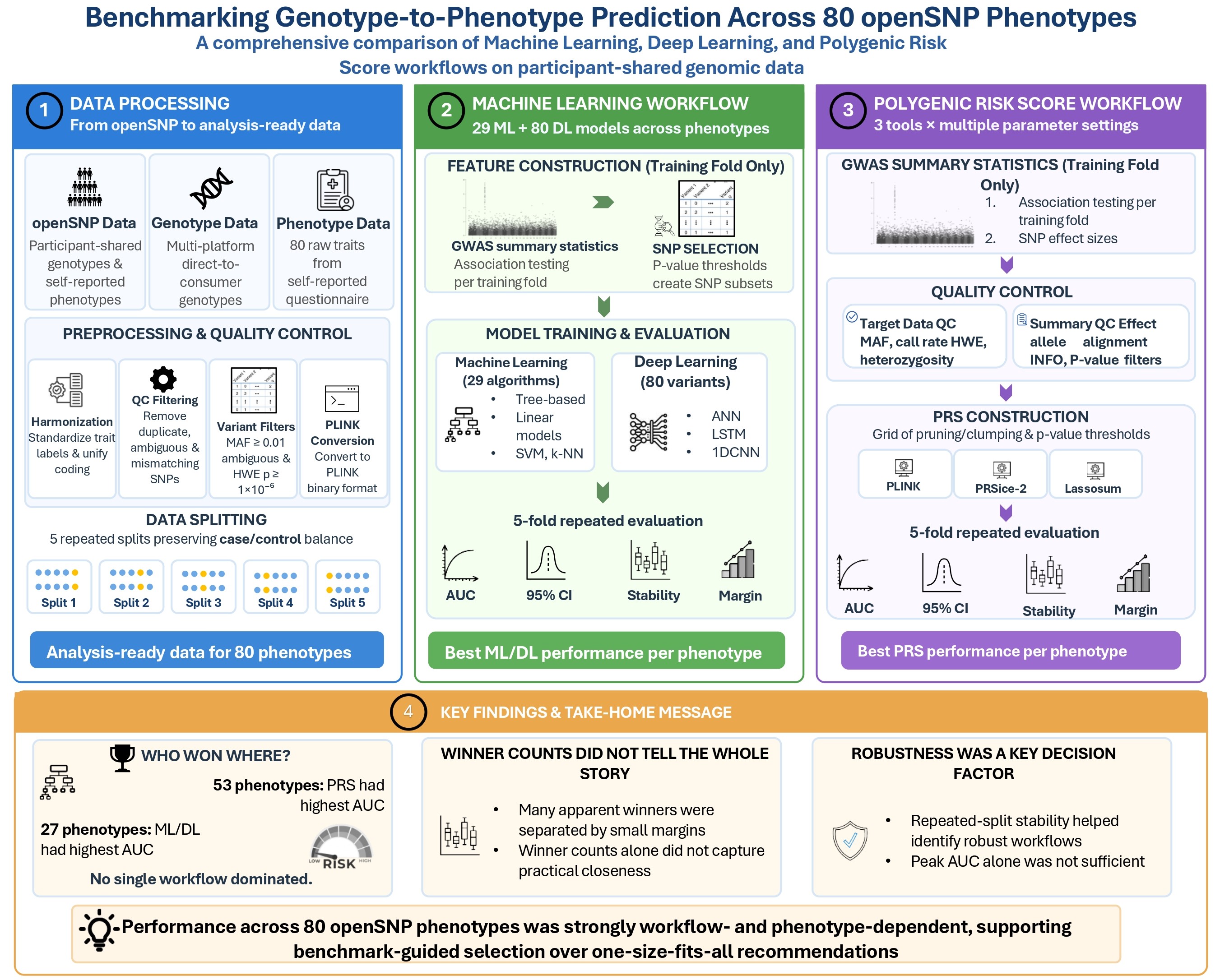}
\caption{\textbf{Overview of the end-to-end benchmark pipeline.} The three workflow arms (Machine Learning, Deep Learning, and Polygenic Risk Score) are shown alongside the SNP subset construction strategy, repeated stratified splits, and evaluation framework applied uniformly across 80 binary phenotypes from openSNP.}
\label{fig:overview}
\end{figure*}

Figure~\ref{fig:overview} summarises the end-to-end benchmark pipeline. Three workflow families were evaluated in parallel across 80 binary phenotypes: 29 ML algorithms, 80 DL model variants, and 3 PRS tools (Plink, PRSice, and Lassosum) across 675 clumping and pruning configurations. For ML and DL workflows, seven SNP subsets were constructed per phenotype using p-value thresholding on training-fold association statistics, corresponding to 50, 100, 200, 500, 1000, 5000, and 10000 SNPs. For PRS workflows, GWAS summary statistics were generated internally from the training fold of each split, as external summary statistics were available for only 30 of 80 phenotypes and carried ancestry mismatch and genotype-rate discrepancies with the openSNP target data. All workflow families were evaluated under five repeated stratified train-test splits (75\%/25\%), and the configuration with the highest mean test AUC across splits was selected per phenotype for reporting. This design constitutes an exploratory benchmark of end-to-end prediction workflows rather than a definitive ranking of algorithms under strict independent test-set evaluation.

\subsection*{No workflow family dominates across phenotypes}
\label{sec:familylevel_results}

\begin{figure*}[!ht]
\includegraphics[width=\textwidth]{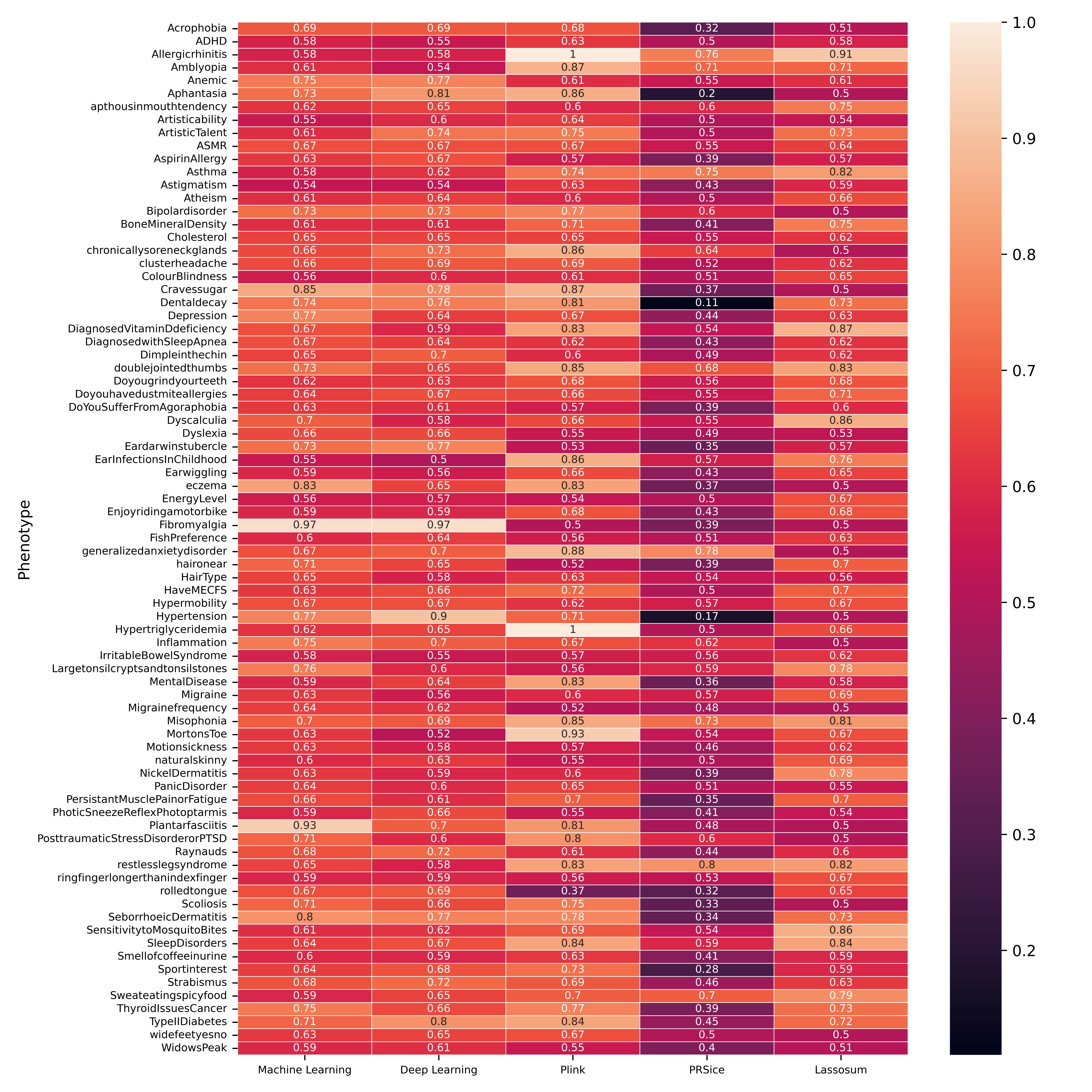}
\caption{\textbf{Heatmap of AUC values across all five workflow families and 80 phenotypes.} Each cell reports the mean AUC as a proportion across five repeated stratified splits. Rows represent phenotypes and columns represent workflow families. Deeper colour intensity indicates lower AUC.}
\label{fig:heatmap}
\end{figure*}

\begin{figure*}[!ht]
\includegraphics[width=\textwidth]{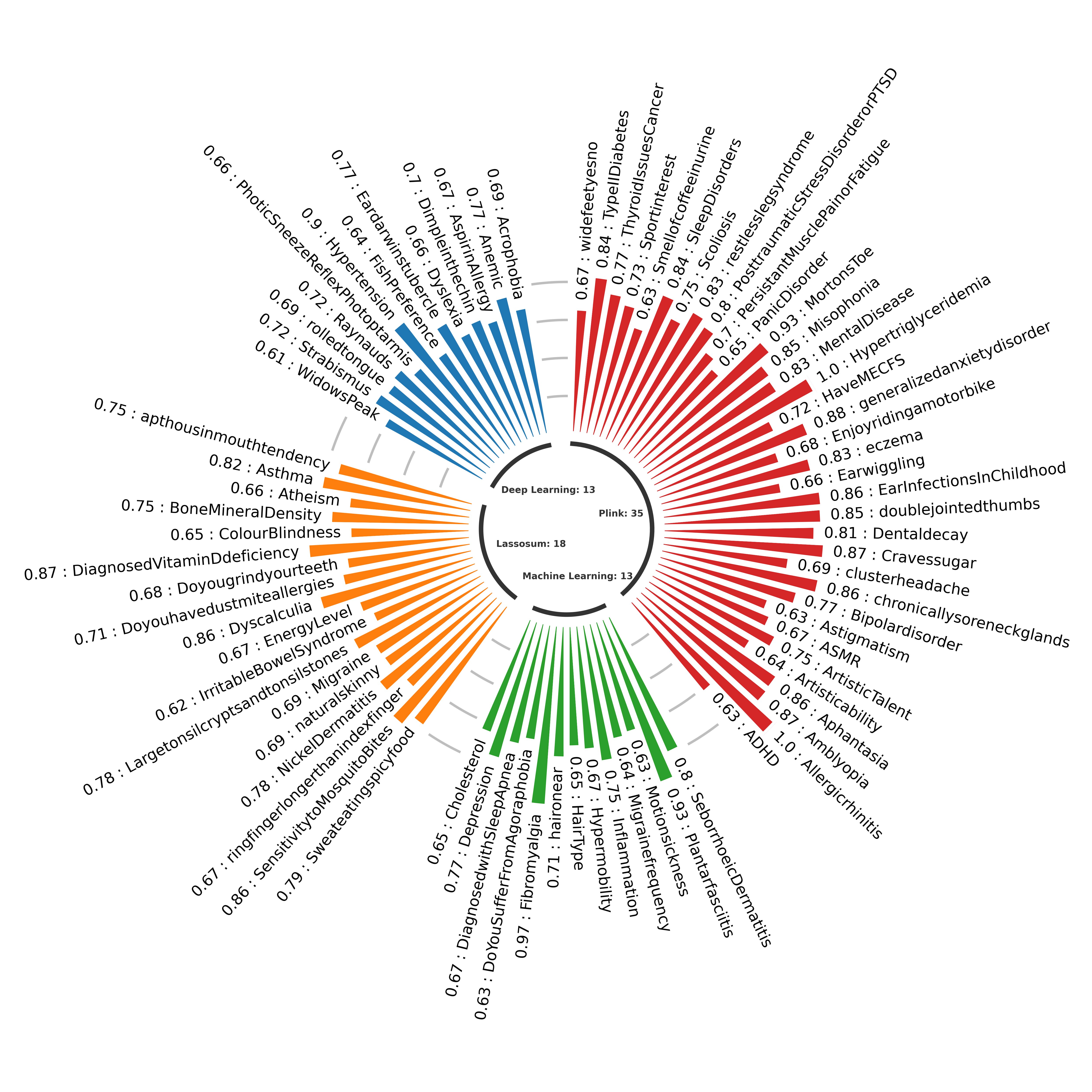}
\caption{\textbf{Circular grouped bar chart of phenotype-level winners across all 80 phenotypes.} Phenotypes are grouped by the workflow family that achieved the highest observed AUC. The AUC value for each phenotype is shown alongside the phenotype name. Colours indicate the winning workflow family: Plink (red), Lassosum (orange), Machine Learning (green), Deep Learning (blue), and PRSice (not shown as no phenotype-level wins were recorded).}
\label{fig:bestperformers}
\end{figure*}

To provide the highest-level comparison before examining individual workflow behaviour, we summarised each workflow family using mean AUC, median AUC, mean CI width, total phenotype-level wins, and decisive wins exceeding 5 AUC points. Figure~\ref{fig:heatmap} shows the full AUC values produced by all five workflow families across the 80 phenotypes, Figure~\ref{fig:bestperformers} groups phenotypes by the workflow family that achieved the highest observed AUC, and Supplementary Table~S2 provides the observed phenotype-level performance values used for these comparisons. Across the 80 phenotypes, Plink achieved the highest mean AUC at 68.15\%, followed by Machine Learning (66.19\%), Deep Learning (65.29\%), Lassosum (63.70\%), and PRSice (48.38\%). Median AUC values showed the same ordering, with Plink at 66.67\%, Deep Learning at 64.60\%, Machine Learning at 64.30\%, Lassosum at 63.37\%, and PRSice at 49.78\%. PRS workflows achieved the highest observed AUC for 53 phenotypes while ML/DL workflows achieved the highest for 27 phenotypes. Among PRS workflows, Plink achieved the highest observed AUC for 35 phenotypes and Lassosum for 18 phenotypes, while PRSice did not achieve the highest observed AUC for any phenotype. Among ML/DL workflows, Deep Learning achieved the highest for 14 phenotypes and Machine Learning for 13 phenotypes. When only decisive wins exceeding 5 AUC points were counted, Plink led with 11 such wins (13.8\%), followed by Lassosum with 10 (12.5\%), Deep Learning with 3 (3.8\%), Machine Learning with 2 (2.5\%), and PRSice with none. When stability was assessed using mean CI width, PRSice showed the least favourable profile at 39.95, while Lassosum and Deep Learning showed the narrowest mean CI widths at 26.71 and 27.50 respectively, followed by Machine Learning at 30.91 and Plink at 31.37. For Fibromyalgia, Plantar Fasciitis, Hypertension, and Eczema, ML/DL workflows produced AUC values at or above 80\%. For Allergicrhinitis, Hypertriglyceridemia, MortonsToe, and generalizedanxietydisorder, PRS workflows produced AUC values at or above 80\%. Table~\ref{tab:familylevel} summarises the family-level performance across all five workflow families.

\subsection*{Winner counts overstate practical differences between workflows}
\label{sec:margin_results}

Because phenotype-wise winner counts can exaggerate the importance of small AUC differences, we complemented the family-level overview with margin-based and paired statistical comparisons. Across the 80 phenotypes with both ML and DL results available, 33 (41.2\%) were practical ties with a margin of at most 5 AUC points between the best ML/DL and best PRS workflow, whereas 30 (37.5\%) showed a clear winner with a margin greater than 10 AUC points. ML and DL showed limited overall separation, while Plink remained frequently competitive against the best ML/DL workflow. Table~\ref{tab:margins} and Supplementary Table~S9 summarise these results.

Paired Wilcoxon signed-rank tests confirmed these patterns (Table~\ref{tab:wilcoxon}). Both ML and DL significantly outperformed PRSice (ML: $W = 209.0$, $p < 0.0001$, $r = 0.76$; DL: $W = 238.0$, $p < 0.0001$, $r = 0.74$), attributable to PRSice's SNP-count normalisation penalty under low genotype rates. Neither ML nor DL significantly outperformed Lassosum ($p > 0.44$, negligible effects). The comparison between DL and Plink was significant but small ($W = 1151.5$, $p = 0.0246$, $r = 0.25$). No significant difference was observed between ML and DL overall ($p = 0.4732$). When the best ML/DL workflow was compared against the best PRS workflow per phenotype, PRS achieved a significantly higher mean AUC (71.34\% versus 68.12\%; $W = 1036.0$, $p = 0.0051$, $r = 0.31$, medium effect), consistent with PRS winning 53 of 80 phenotypes.

\subsection*{Workflow-specific diagnostics reveal distinct instability and failure modes}
\label{sec:workflowbehaviour_results}

To understand the family-level patterns, we examined configuration dependence, instability, and collapse within each workflow family.

\begin{figure*}[!ht]
\centering
\includegraphics[width=\textwidth]{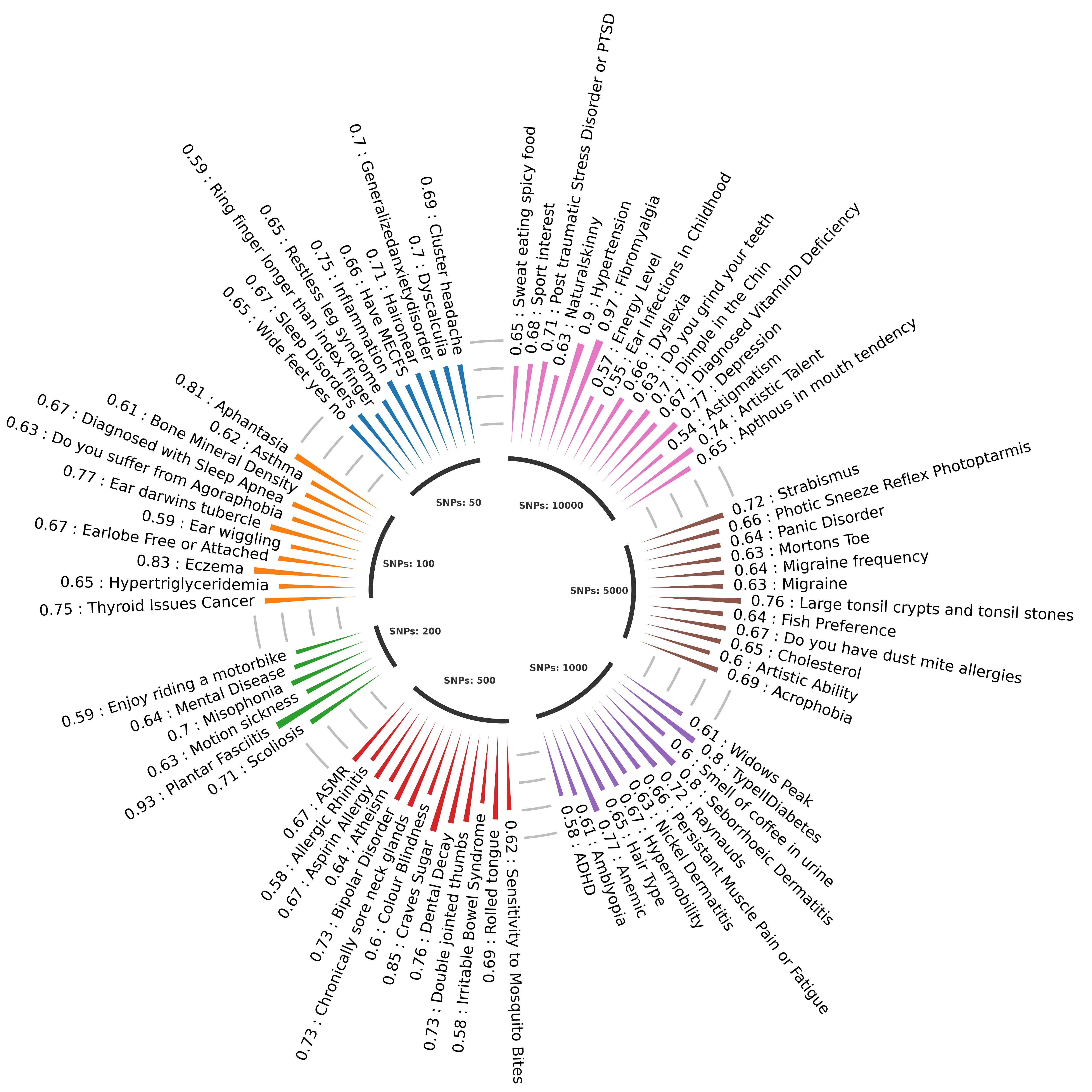}
\caption{\textbf{AUC values across phenotypes grouped by the SNP subset yielding the highest observed ML/DL performance.}}
\label{groupbysnps}
\end{figure*}

Figure~\ref{groupbysnps} shows ML and DL AUC values across all 80 phenotypes grouped by the SNP subset yielding the highest observed performance. The optimal SNP subset varied substantially across phenotypes, indicating phenotype-specific signal structure. XGBoost was the most frequently selected ML algorithm, whereas ANN was the most frequently selected DL architecture (Table~\ref{tab:dl_algorithm_counts}). The most common DL hyperparameter combination was dropout 0.2, Adam, batch size 1, and 50 epochs (Table~\ref{tab:dl_hyperparameter_counts}). ML and DL AUC values spanned broad ranges, with wider confidence intervals in several small phenotypes and one DL collapse case for Post-traumatic Stress Disorder. Full per-phenotype results are provided in Supplementary Tables~S4 and S5.

For the PRS workflows, we applied 675 parameter combinations to each tool. Table~\ref{tab:plink_params} shows the most frequently selected Plink parameter combinations among phenotype-level winners. Plink winner AUC values ranged from 50.0\% to 100.0\%; the 100.0\% values observed for Allergicrhinitis and Hypertriglyceridemia are likely artefacts of severe class imbalance combined with small sample sizes and should be interpreted cautiously. Plink achieved the highest overall AUC for 35 phenotypes (43.8\%) and fell within 5 AUC points of the top-performing workflow for 44 phenotypes (55.0\%), indicating that it was frequently either the strongest or a near-competitive workflow. Plink exhibited no model collapses and no zero-SD cases.

For PRSice, a single parameter combination (200-50-0.1-200-1-0.1) yielded the highest AUC for 72 of 79 phenotypes (91.1\%), indicating near-complete concentration of phenotype-level winners within one configuration. PRSice produced the weakest overall performance profile, yielding the lowest AUC among all five workflow families for 63 phenotypes (78.8\%). Its poor performance is most likely explained by its SNP-count normalisation formula, which penalises datasets with low genotype rates such as openSNP. PRSice also produced the most unstable results overall, with 36 phenotypes showing CI width above 40 percentage points and 5 phenotypes producing negative CI lower bounds under symmetric $t$-based interval estimation. 

Table~\ref{tab:lassosum_params} shows the most frequently selected Lassosum parameter combinations among phenotype-level winners. A notable failure mode in Lassosum was collapse to non-informative prediction, with 13 phenotypes producing an AUC of exactly 50.0\% and zero standard deviation across all splits. Lassosum achieved a mean peak AUC of 64.37\% with a mean CI width of 26.71, but these collapse cases should be interpreted as prediction failure rather than genuine stability.

To assess reproducibility, we computed the 95\% CI width per tool and phenotype across the five repeated stratified splits (Table~\ref{tab:stability}). Lassosum and Deep Learning showed the narrowest mean CI widths, whereas PRSice showed the widest and therefore the least favourable stability profile. A Kruskal-Wallis test on CI widths across workflow families was significant ($H = 22.01$, $p = 0.0002$), and post-hoc Mann-Whitney U tests showed that PRSice produced significantly wider CI than Machine Learning, Deep Learning, Plink, and Lassosum. Detailed per-phenotype stability and PRS results are provided in Supplementary Tables~S3--S8.

\subsection*{Winning configurations are phenotype-specific for ML/DL but concentrated for PRS workflows}
\label{sec:hyperconc_results}

To assess whether phenotype-level winners were concentrated within a small number of recurring configurations, we ranked the highest-performing configuration for each phenotype within each workflow family and summarised the frequency of the most common settings. ML and DL winners were highly dispersed, with 79 distinct winning configurations across 80 phenotypes in both workflow families and only minimal concentration in the top-ranked settings. In contrast, PRS workflows showed greater concentration, especially PRSice, for which a single parameter combination accounted for 72 of 79 phenotypes (91.1\%). Plink and Lassosum showed intermediate concentration patterns. Table~\ref{tab:hyperconc} summarises these results. Overall, ML and DL winners were strongly phenotype-specific, whereas PRS workflows, particularly PRSice, were more concentrated within a small number of recurring parameter settings.

\subsection*{High peak performance is enriched in smaller phenotypes and requires cautious interpretation}
\label{sec:samplesize_results}

To determine whether phenotype sample size influenced the benchmark results, we correlated total sample size with the highest observed phenotype-level AUC values across the 79 phenotypes with sample-size information. Sample size was negatively correlated with all three summary performance measures, indicating that higher peak AUC values were more often observed in smaller phenotypes, whereas larger phenotypes tended to yield more modest performance. In contrast, class imbalance showed only weak associations with CI width across workflow families and did not strongly explain fold-to-fold instability.

Stratification by sample size confirmed that peak performance was concentrated in smaller phenotypes, with mean best overall AUC declining from small to medium to large phenotypes. PRSice performed poorly across all strata. These findings argue against interpreting high performance in small cohorts as evidence of deployment-ready prediction.

\section{Discussion}

This study provides a large-scale benchmark of genotype-to-phenotype prediction workflows in participant-shared genomic data and shows that no workflow family is uniformly superior across traits. Although polygenic score workflows more often achieved the highest observed discrimination, the practical separation between workflow families was frequently modest, with many phenotype-level comparisons falling within small margins. This is an important point: winner counts alone can overstate meaningful differences, particularly in noisy, phenotype-heterogeneous settings. In this benchmark, the most defensible conclusion is therefore not that one modelling family dominates, but that performance is strongly phenotype-dependent and must be interpreted together with uncertainty, stability, and workflow-specific failure modes.

One of the clearest findings is that benchmark interpretation changes materially when stability is considered alongside mean discrimination. PRSice showed the weakest overall profile, with low mean performance and broad confidence intervals, whereas lassosum displayed a different failure mode, collapsing to non-informative predictions for a subset of phenotypes. Plink provided the strongest overall balance among the polygenic score tools, while machine learning and deep learning were often competitive but highly phenotype-specific in both configuration and SNP-subset dependence. These results reinforce the idea that end-to-end workflow behaviour reflects not only model class, but also the interaction between data properties, preprocessing choices, and score-construction assumptions.

Our results also highlight a broader methodological caution for genomic prediction benchmarks in limited-data settings. Peak discrimination was negatively correlated with phenotype sample size, indicating that the most striking values were concentrated in smaller phenotypes, where performance estimates are inherently less stable and more susceptible to chance separation. This does not invalidate the benchmark, but it does argue strongly against reading high phenotype-level performance in small cohorts as evidence of deployment-ready prediction. In the same way, the internal generation of summary statistics for polygenic score construction means that the present results should be interpreted as a benchmark of small-sample score-construction workflows, not as a surrogate for classical large-discovery polygenic risk studies.

A further insight concerns the biological interpretation of model classes. Artificial neural networks were the most frequently competitive deep-learning architecture, whereas recurrent variants were occasionally effective but are difficult to interpret mechanistically in this setting because SNP order does not have the same temporal semantics as sequence data in language or time-series applications. Their apparent utility is more plausibly viewed as a property of input representation and correlation structure than as evidence for biologically meaningful sequence modelling. More generally, the dispersion of winning machine-learning and deep-learning configurations across phenotypes suggests that flexible phenotype-specific tuning can be valuable, but also that broad claims about algorithm superiority should be treated cautiously when preprocessing and model search spaces differ.

Several limitations remain important. The openSNP cohort is heterogeneous, self-reported, and predominantly of European ancestry; covariates are limited; and the repeated-split design used here is exploratory rather than fully nested. These factors constrain generalisability and mean that the reported performance values should be interpreted as comparative evidence about workflow behaviour under a consistent benchmark, not as final external estimates. Even so, that is precisely where the value of the study lies. By treating openSNP as a realistic stress-test environment, the manuscript provides practical guidance for researchers working with small, heterogeneous genomic datasets and offers a framework that can be extended as larger, better-curated, and more diverse public cohorts become available.

\section{Materials and Methods}
\subsection{Dataset}
\label{sec:dataset}

The openSNP database is a crowdsourced platform that enables individuals to publicly share genotype data obtained from direct-to-consumer (DTC) genetic testing companies including 23andMe, Family Tree DNA, and AncestryDNA \cite{Greshake2014}. At the time of analysis, the dataset contained 6,401 genotype files across 668 phenotypes. The majority of respondents came from the USA (60.33\%), followed by Canada (5.17\%) and the UK (4.61\%). An inherent limitation of this resource is the absence of standardised demographic and population information, as the data originate from multiple DTC providers using different genotyping platforms, chip designs, and missingness profiles.

We considered binary phenotypes from the openSNP dataset. The 83 candidate phenotypes varied substantially in sample size and class balance (Supplementary Table~S1). Total sample sizes ranged from 2 (Nicotinedependence) to 291 (ColourBlindness), with a median of 55 samples. Of the 83 phenotypes, 18 (21.7\%) had fewer than 30 total samples, representing a high-risk group for unreliable model estimates. Class imbalance was also prevalent: 39.8\% of phenotypes were balanced (imbalance ratio $\leq$1.5), 34.9\% were mildly imbalanced (ratio 1.5--3.0), 13.3\% moderately imbalanced (ratio 3.0--5.0), and 12.0\% severely imbalanced (ratio $>$5.0). Severely imbalanced phenotypes included Colour Blindness (ratio = 15.17), Allergicrhinitis (11.00), and Aspirin Allergy (7.72). These characteristics motivated the use of AUC as the primary evaluation metric, as it is robust to class imbalance, and informed the cautious interpretation of results for small-sample phenotypes throughout this study.

\subsection{Phenotype Harmonisation and Genotype Preprocessing}
\label{sec:preprocessing}

Phenotype label responses by participants were inconsistent across the dataset, requiring manual cleaning. For example, right-handedness was recorded as ``right-handed'', ``right'', ``Right'', and ``R'' by different participants. Each phenotype value was mapped to one of three categories: Case (presence of the phenotype value), Control (absence), or Unknown (ambiguous or unclassifiable response). Values labelled Unknown were excluded from analysis. This process is referred to as phenotype cleaning or transformation \cite{AlvarezPrado2019}, and it required approximately three to five days of manual effort. The transformation files are available on GitHub (\texttt{preTransform.csv}, \texttt{postTransform.csv}, \texttt{AmbiguousPhenotypes.pdf}). 

Genotype files were provided in 23andMe, DecodeMe, and FamilyTreeDNA formats. All files were converted to the 23andMe format and subsequently to Plink binary format (bed, bim, fam) \cite{9802470}. After conversion, 101 phenotypes were retained for quality control. The following filters were applied using PLINK \cite{Purcell2007}: minor allele frequency $\geq$ 0.01, Hardy--Weinberg equilibrium $p \geq 10^{-6}$, genotype missingness rate $\leq$ 0.01, and individual missingness rate $\leq$ 0.7. Duplicate individuals and duplicate SNPs were removed at this stage. After quality control, 80 phenotypes with sufficient sample sizes and genotype data quality were retained for downstream analysis. A complete list of retained phenotypes is available on GitHub (\texttt{Analysis3.pdf}).

\subsection{Population Structure Analysis}
\label{sec:popstruct}

To characterise the ancestry composition of the openSNP cohort, we performed principal component analysis (PCA) by projecting the openSNP genotype data onto the 1000 Genomes Phase 3 reference panel. All 83 per-phenotype genotype files were merged into a single dataset retaining 1,188 unique individuals after deduplication. Quality control was applied with minor allele frequency $>$ 0.01, genotype missingness $<$ 0.05, and Hardy--Weinberg equilibrium $p > 10^{-6}$. Variant identifiers were harmonised between datasets using chromosome and base-pair position (CHR:BP) matching to account for differences in SNP naming conventions between DTC genotyping platforms and the reference panel. Common SNPs were extracted, followed by linkage disequilibrium pruning (window = 200, step = 50, $r^2$ threshold = 0.2). The merged dataset was projected onto the first ten principal components using PLINK, and each openSNP sample was assigned to its nearest 1000 Genomes superpopulation centroid in PC1--PC6 space.

\subsection{Evaluation Framework}
\label{sec:evalframework}
Data were split into training (75\%) and test (25\%) sets using five repeated stratified random splits generated by \texttt{StratifiedShuffleSplit}. Workflow configurations were explored for ML, DL, and PRS tools, and the configuration with the highest mean test AUC across the five splits was selected for reporting. Because configuration selection and performance estimation were conducted within the same repeated-split framework, the reported AUC values should be interpreted as exploratory resampling results rather than as estimates from a strictly nested independent test-set design. Performance was quantified using the area under the receiver operating characteristic curve (AUC), which is robust to class imbalance \cite{Khera2018,Wray2010,Medvedev2022}. To quantify performance variability across repeated splits, we computed 95\% confidence intervals (CIs) for the mean AUC of each workflow using a $t$-distribution with four degrees of freedom:

\begin{equation}
CI = \bar{AUC} \pm \left( t_{0.975, 4} \times \frac{SD}{\sqrt{5}} \right)
\label{eq:ci}
\end{equation}

where $t_{0.975, 4} = 2.776$, $SD$ is the standard deviation of AUC across the five splits, and $\sqrt{5}$ is the number of repeated splits. This symmetric interval estimator can occasionally produce bounds outside the $[0, 1]$ range for bounded metrics under small sample sizes; such cases are flagged diagnostically in the stability analysis and do not reflect meaningful biological signal. A supplementary sensitivity analysis confirms that the family-level conclusions reported here are robust to this interval choice (Supplementary Table~S10).

\subsection{Machine Learning Workflow}
\label{sec:mlworkflow}

Genotype data typically consist of millions of SNPs, necessitating dimensionality reduction before modelling. We applied p-value thresholding using training-fold GWAS summary statistics \cite{Adam2021,Chen2021,al2021deep}. The training data for each fold were subjected to Fisher's exact test for allelic association using \texttt{plink --assoc} to generate per-SNP p-values. Seven SNP subsets were then extracted from both the training and test data by linearly increasing the p-value threshold from zero until the target number of SNPs was reached: 50, 100, 200, 500, 1000, 5000, and 10000 SNPs. The encoded data were saved in raw spreadsheet format and passed to ML and DL models. The p-value threshold required to extract a given number of SNPs varies across phenotypes depending on the underlying association signal \cite{Silva2022,Medvedev2022}.

\subsubsection{Machine Learning Models}
\label{sec:mlmodels}

We used 29 classical ML algorithms with default parameters from the \texttt{scikit-learn} library \cite{scikit-learn}. The algorithms included tree-based classifiers \cite{asd} such as AdaBoost, XGBoost \cite{Chen:2016:XST:2939672.2939785}, Random Forest, and Gradient Boosting, as well as Stochastic Gradient Descent (SGD), Multi-layer Perceptron (MLP), Support Vector Classifier, and additional classifiers. The complete list of algorithms and hyperparameters is available on GitHub (\texttt{MachineLearningAlgorithms.txt}).

\subsubsection{Deep Learning Models}
\label{sec:dlmodels}

We used four base DL architectures: Artificial Neural Network (ANN) \cite{mcculloch1943logical}, Gated Recurrent Unit (GRU) \cite{8053243}, Long Short-Term Memory (LSTM) \cite{Hochreiter1997}, and Bidirectional LSTM (BILSTM) \cite{Schuster1997}. Recurrent neural network variants have previously been applied to genotype-phenotype prediction \cite{Pouladi2015,Srinivasu2022}. It should be noted that the biological rationale for recurrent architectures on SNP sequences is limited, as genomic SNP order does not directly correspond to temporal sequence structure, and any apparent gains from recurrent models may reflect linkage disequilibrium structure or input encoding rather than biologically meaningful long-range dependencies. All architectures used five layers \cite{Uzair2020,PrezEnciso2019}, with neurons per layer defined as $128 + \sqrt{S}$, $64 + \sqrt[4]{S}$, $32 + \sqrt[8]{S}$, $16 + \sqrt[16]{S}$, and 1, where $S$ is the number of input SNPs. Six additional stacked architectures combined GRU, LSTM, and BILSTM layers. Hyperparameters spanned dropout (0.2, 0.5), batch size (1, 5), and epochs (50, 200) with Adam as the optimiser, yielding 80 DL variants in total (\texttt{DeepLearningAlgorithms.txt}).

\subsection{Polygenic Risk Score Workflow}
\label{sec:prsworkflow}

PRS tools require a base GWAS summary-statistics file and a target genotype file. Because suitable external GWAS summary statistics were available for only 30 of the 80 phenotypes and often differed in ancestry composition and genotype rate from the openSNP target data, we generated GWAS summary statistics internally from the training fold of each split using Fisher's exact test via \texttt{plink --assoc}. This design should therefore be interpreted as an internal small-sample score-construction benchmark rather than a classical PRS evaluation based on large external discovery cohorts. The following columns were generated for the GWAS summary file: chromosome (CHR), base pair position (BP), SNP identifier (SNP), p-value (P), odds ratio (OR), standard error (SE), reference allele (A1), alternative allele (A2), minor allele frequency (MAF), and total sample size (N). Chromosomes, base pair positions, SNP identifiers, p-values, and odds ratios were generated using Plink; standard errors were derived from P and OR; and reference allele, alternative allele, MAF, and N were generated using SNPTEST. The number of samples available for most phenotypes is relatively low but still meets minimum criteria for score construction \cite{Crossa1989,Qu2019}.

\subsubsection{Quality Control for PRS}
\label{sec:prsqc}

Quality control was applied separately to the GWAS summary file and the target genotype data. For the GWAS file, duplicate, ambiguous, and mismatched SNPs were removed, and SNPs with imputation information score $<$ 0.8 or minor allele frequency $<$ 0.01 were excluded. For the target genotype data, duplicate, ambiguous, and mismatched SNPs were removed, and SNPs with genotype rate $<$ 0.01, minor allele frequency $<$ 0.01, or Hardy--Weinberg equilibrium $p < 10^{-6}$ were excluded. Individuals with a per-SNP missing rate exceeding 0.7 were removed. Downstream analysis was performed on the common SNPs between the GWAS and target files \cite{Anderson2010,Laurie2010}. Pruning and clumping were applied to the target data across the parameter grid shown in Table~\ref{PRSparameters}, yielding $3 \times 3 \times 3 \times 1 \times 5 \times 5 = 675$ parameter combinations per tool. The complete parameter list is available on GitHub (\texttt{Plink\_PRSice\_Lassosum\_Parameters.txt}). Rather than converting PRS scores to binary case/control labels prior to evaluation, the continuous PRS scores were passed directly to the AUC calculation, preserving the full ranking information contained in the score distribution \cite{Choi2020}.

\subsubsection{PRS Tools}
\label{sec:prstools}

We evaluated three PRS tools: Plink \cite{Purcell2007}, PRSice2 \cite{Euesden2014}, and Lassosum \cite{Mak2017}. LDpred2 was not included because it requires SNPs from the 1000 Genomes or HapMap3 reference panels and relies on heritability estimation as a prerequisite, which is not appropriate for phenotypes lacking clear genetic correlations \cite{Choi2020}. Plink computes PRS by summing the products of effect sizes and genotype dosages across variants:

\begin{equation}
PRS_{\text{Plink}} = \sum_{i=1}^{n} \left( \beta_i \times G_i \right)
\label{Plinkequation}
\end{equation}

where $\beta_i$ is the effect size of the $i$th SNP from the GWAS summary file, $G_i \in \{0, 1, 2\}$ is the genotype dosage, and $n$ is the total number of SNPs. PRSice2 computes PRS using the same summation but divides by the total number of SNPs:

\begin{equation}
PRS_{\text{PRSice}} = \sum_i \frac{S_i \times G_i}{M}
\label{prsiceequation}
\end{equation}

where $S_i$ and $G_i$ are the effect size and genotype value of the $i$th SNP, and $M$ is the total number of SNPs. This normalisation penalises datasets with low genotype rates, as the denominator includes all SNPs regardless of whether they have observed genotypes. Lassosum applies LASSO regression to the GWAS summary statistics to shrink effect sizes and compute a regularised PRS:

\begin{equation}
PRS_{\text{Lassosum}} = \sum_{i=1}^{p} X_i \beta_i
\label{lassosumequation}
\end{equation}

where the coefficients $\beta_i$ are estimated by minimising:

\begin{equation}
\text{Lasso}(\beta) = \left\| y - \sum_{i=1}^{p} X_i \beta_i \right\|_2 + 2\lambda \left\| \sum_{i=1}^{p} \beta_i \right\|_1
\label{objectivelasso}
\end{equation}

where $\lambda$ is the regularisation parameter that shrinks small coefficients towards zero, $y$ is the phenotype vector, and $p$ is the number of SNPs.

\subsection{Statistical Analyses}
\label{sec:statsanalyses}

\subsubsection{Stability and Margin Analysis}
\label{sec:stability}

To evaluate the reproducibility of workflow performance beyond point estimates, CI width was computed per tool and phenotype across the five repeated splits and used as a measure of stability, where narrower intervals indicate more consistent performance. ML and DL results were available for all 80 phenotypes. For Plink, PRSice, and Lassosum, one phenotype (EarlobeFreeorattached) did not have a complete set of usable PRS result files and was therefore excluded from PRS stability analyses, yielding 79 phenotypes for those analyses. Each phenotype-tool result was assigned one or more diagnostic flags: \texttt{MODEL\_COLLAPSE} when AUC = 50.0\% and SD = 0 across all splits, indicating non-informative predictions; \texttt{ZERO\_SD} when SD = 0 but AUC $\neq$ 50.0\%; \texttt{WIDE\_CI} when CI width exceeded 40 percentage points; and \texttt{NEGATIVE\_CI} when the lower CI bound fell below 0, a mathematical artefact of symmetric $t$-based interval estimation applied to a bounded metric under small-sample conditions. The per-phenotype stability summary is reported in Supplementary Table~S3 and the full workflow-specific stability results in Supplementary Tables~S4--S8. The combined per-phenotype risk flag summary is reported in 
Supplementary Table~S11. A Kruskal-Wallis test was used to assess whether CI widths differed significantly across workflow families, and pairwise Mann-Whitney U tests were used for post-hoc comparisons. A phenotype-level margin analysis was additionally performed to quantify how large the performance differences were between workflow families, rather than relying only on winner counts. Absolute AUC differences were grouped into five bins: $<1$ (negligible), 1--3 (trivial), 3--5 (small), 5--10 (moderate), and $>10$ (meaningful) AUC points. The full per-phenotype margin results are reported in Supplementary Table~S9.

\subsubsection{Paired Statistical Comparison}
\label{sec:wilcoxon}

Paired Wilcoxon signed-rank tests were applied to phenotype-level AUC values across the 80 phenotypes with both ML and DL results available. Eight comparisons were evaluated: ML versus PRSice, DL versus PRSice, ML versus Lassosum, DL versus Lassosum, ML versus Plink, DL versus Plink, ML versus DL, and best ML/DL versus best PRS, where best ML/DL = $\max(\text{ML}, \text{DL})$ and best PRS = $\max(\text{Plink}, \text{PRSice}, \text{Lassosum})$. Effect size was computed as $r = |z| / \sqrt{n_{\text{nonzero}}}$, where $z$ is the normal approximation of the Wilcoxon statistic and $n_{\text{nonzero}}$ is the number of non-zero paired differences. Effect sizes were interpreted as negligible ($r < 0.10$), small ($0.10 \leq r < 0.30$), medium ($0.30 \leq r < 0.50$), and large ($r \geq 0.50$).

\subsubsection{Hyperparameter Concentration, Sample Size, and Robustness Analyses}
\label{sec:additional}

Winning configurations were ranked by frequency within each workflow family to assess hyperparameter concentration. ML and DL configurations were defined by algorithm/architecture and SNP count; PRS configurations by the full pruning and clumping parameter string. Pearson and Spearman correlations were computed between sample size and peak AUC, and between imbalance ratio and CI width. Phenotypes were stratified into small ($<30$), medium (30--99), and large ($\geq100$) sample groups. A sensitivity analysis excluding phenotypes flagged for instability confirmed the robustness of family-level conclusions (Supplementary Table~S10).
  
\section*{Ethics approval and consent to participate}

This study used publicly available, participant-shared genotype and phenotype data from openSNP. No new human participants were recruited and no new primary human-subject data were generated. All analyses were performed on publicly available de-identified data. Institutional ethics review was not required for this secondary analysis of public data.

\section*{Acknowledgements}

D.B.A. is supported by an NHMRC Investigator Grant (GNT2041888).

\section*{Funding}
M.M. acknowledges scholarship support as a PhD student. D.B.A. is supported by an NHMRC Investigator Grant (GRNT2041888). This research was also supported by the NVIDIA Academic Grant Program.

\section*{Competing Interest}
The authors declare that they have no competing interests.

\section*{CRediT Author Statement}
Muhammad Muneeb: Conceptualization, Formal analysis, Investigation, Methodology, Software, Visualization, Writing -- original draft. David B. Ascher: Conceptualization, Funding acquisition, Resources, Supervision, Writing -- review \& editing. YooChan Myung: Formal analysis, Writing -- review \& editing. Samuel F. Feng: Methodology, Writing -- review \& editing. Andreas Henschel: Methodology, Writing -- review \& editing. All authors have read and approved the final manuscript.

\section*{ORCID}
0000-0002-6506-4430 (Muhammad Muneeb)\newline
0000-0003-2948-2413 (David B. Ascher)\newline
0000-0002-6763-9808 (YooChan Myung)\newline
0000-0001-9046-2601 (Samuel F. Feng)\newline
0000-0003-1386-5372 (Andreas Henschel)

\section*{Availability of Data and Materials}
The genotype and phenotype data analysed in this study are publicly available from openSNP (\url{https://opensnp.org/}). The code and supporting materials are available at \url{https://github.com/MuhammadMuneeb007/Benchmarking-80-OpenSNP-phenotypes-using-deep-learning-algorithms-and-polygenic-risk-scores-tools/}.

\section*{Supplementary Material}
Supplementary Tables S1--S11 are provided with the submission. Supplementary material will be available at \textit{Genomics, Proteomics \& Bioinformatics} online.

\bibliographystyle{oup-plain}
\bibliography{reference}
 
\clearpage
\section*{Figures}

\clearpage

\section*{Tables}

\begin{table*}[!ht]
\centering
\caption{\textbf{PRS parameter grid used for pruning and clumping.} Clumping $p1$ denotes the significance threshold for index SNPs, $r^2$ the linkage disequilibrium threshold, and kb the physical distance threshold. The full grid yielded 675 $(3 \times 3 \times 3 \times 1 \times 5 \times 5)$ parameter combinations per tool.}
\label{PRSparameters}
\begin{tabular}{ll}
\hline
\textbf{Parameter} & \textbf{Values} \\
\hline
Pruning window size & 200, 500, 1000 \\
Pruning window shift size & 50, 100, 150 \\
Pruning LD threshold & 0.1, 0.3, 0.5 \\
Clumping $p1$ & 1 \\
Clumping $r^2$ & 0.1, 0.3, 0.5, 0.7, 0.9 \\
Clumping kb & 200, 400, 600, 800, 1000 \\
\hline
\end{tabular}
\end{table*}

\begin{table*}[!ht]
\centering
\caption{\textbf{Deep learning algorithms ranked by phenotype-level winner count.} The first column lists the DL architecture and the second column shows the number of phenotypes for which that architecture yielded the highest observed AUC.}
\label{tab:dl_algorithm_counts}
\begin{tabular}{lr}
\hline
\textbf{Deep Learning Algorithm} & \textbf{Phenotype Count} \\
\hline
ANN & 26 \\
Stack-LSTM-BILSTM & 10 \\
Stack-BILSTM-GRU & 8 \\
Stack-GRU-LSTM & 7 \\
GRU & 7 \\
BILSTM & 6 \\
Stack-BILSTM-LSTM & 5 \\
Stack-LSTM-GRU & 5 \\
Stack-GRU-BILSTM & 5 \\
LSTM & 1 \\
\hline
\end{tabular}
\end{table*}

\begin{table*}[!ht]
\centering
\caption{\textbf{Deep learning hyperparameter combinations ranked by phenotype-level winner count.} Each row shows a unique combination of dropout, optimiser, batch size, and epoch count, along with the number of phenotypes for which that combination yielded the highest observed AUC.}
\label{tab:dl_hyperparameter_counts}
\begin{tabular}{llllr}
\hline
\textbf{Dropout} & \textbf{Optimizer} & \textbf{Batch Size} & \textbf{Epochs} & \textbf{Phenotype Count} \\
\hline
0.2 & Adam & 1 & 50 & 23 \\
0.2 & Adam & 5 & 50 & 18 \\
0.5 & Adam & 5 & 50 & 16 \\
0.5 & Adam & 1 & 50 & 15 \\
0.2 & Adam & 5 & 200 & 4 \\
0.5 & Adam & 1 & 200 & 2 \\
0.2 & Adam & 1 & 200 & 1 \\
0.5 & Adam & 5 & 200 & 1 \\
\hline
\end{tabular}
\end{table*}

\begin{table*}[!ht]
\centering
\caption{\textbf{Most frequent Plink parameter combinations among phenotype-level winners.} Three configurations tied at 7 phenotypes each. Parameters are: pruning window size, pruning shift, pruning LD threshold, clumping KB, clumping P1, and clumping R2.}
\label{tab:plink_params}
\begin{tabular}{llllllr}
\hline
\textbf{Window} & \textbf{Shift} & \textbf{LD} & \textbf{KB} & \textbf{P1} & \textbf{R2} & \textbf{Count} \\
\hline
200 & 50 & 0.1 & 1000 & 1 & 0.1   & 7 \\
200 & 50 & 0.1 & 200  & 1 & 0.775 & 7 \\
200 & 50 & 0.1 & 200  & 1 & 0.1   & 7 \\
200 & 50 & 0.1 & 600  & 1 & 0.1   & 4 \\
200 & 50 & 0.1 & 200  & 1 & 0.325 & 4 \\
200 & 50 & 0.1 & 400  & 1 & 0.325 & 3 \\
200 & 50 & 0.1 & 800  & 1 & 0.1   & 3 \\
200 & 50 & 0.1 & 200  & 1 & 0.55  & 3 \\
\hline
\end{tabular}
\end{table*}

\begin{table*}[!ht]
\centering
\caption{\textbf{Most frequent Lassosum parameter combinations among phenotype-level winners.} Parameters are: pruning window size, pruning shift, pruning LD threshold, clumping KB, clumping P1, and clumping R2.}
\label{tab:lassosum_params}
\begin{tabular}{llllllr}
\hline
\textbf{Window} & \textbf{Shift} & \textbf{LD} & \textbf{KB} & \textbf{P1} & \textbf{R2} & \textbf{Count} \\
\hline
200  & 50  & 0.1 & 200 & 1 & 0.1 & 14 \\
500  & 150 & 0.1 & 200 & 1 & 0.1 & 6  \\
200  & 150 & 0.3 & 200 & 1 & 0.1 & 5  \\
1000 & 50  & 0.3 & 200 & 1 & 0.1 & 5  \\
1000 & 50  & 0.1 & 200 & 1 & 0.1 & 4  \\
500  & 50  & 0.1 & 200 & 1 & 0.1 & 4  \\
\hline
\end{tabular}
\end{table*}

\begin{table*}[!ht]
\centering
\caption{\textbf{Hyperparameter concentration among phenotype-level winners.} For each workflow family, the table reports the number of phenotypes analysed, the number of distinct winning configurations observed, the most frequent winning configuration, and the proportion of phenotype winners explained by the top-ranked and top three configurations combined. PRS tools cover 79 phenotypes after exclusion of EarlobeFreeorattached.}
\label{tab:hyperconc}
\resizebox{\textwidth}{!}{%
\begin{tabular}{lrrp{6.5cm}rrrr}
\hline
\textbf{Tool} & \textbf{Phenotypes} & \textbf{Unique Configs} & \textbf{Top 1 Configuration} & \textbf{Top 1 Count} & \textbf{Top 1 \%} & \textbf{Top 3 Count} & \textbf{Top 3 \%} \\
\hline
Machine Learning & 80 & 79 & XGBoost, gblinear, binary:hinge, SNPs=1001 & 2 & 2.5 & 4 & 5.0 \\
Deep Learning & 80 & 79 & ANN, Dropout=0.2, Adam, Batch=5, Epochs=50, SNPs=101 & 2 & 2.5 & 4 & 5.0 \\
Plink & 79 & 44 & 200-50-0.1-1000-1-0.1 (tied with two others at 7) & 7 & 8.9 & 21 & 26.6 \\
PRSice & 79 & 8 & 200-50-0.1-200-1-0.1 & 72 & 91.1 & 74 & 93.7 \\
Lassosum & 79 & 26 & 200-50-0.1-200-1-0.1 & 14 & 17.7 & 25 & 31.6 \\
\hline
\end{tabular}%
}
\end{table*}

\begin{table*}[!ht]
\centering
\caption{\textbf{Stability of workflow families across phenotypes.} N = number of phenotypes evaluated; ML and DL cover 80 phenotypes, PRS tools cover 79 phenotypes after exclusion of EarlobeFreeorattached. Mean CI Width = mean 95\% confidence interval width computed as $2 \times t_{0.975,4} \times SD/\sqrt{5}$. N CI$<$20 = number of phenotypes with CI width below 20 percentage points. N Collapse = phenotypes with AUC = 50.0\% and SD = 0 (model collapse). N Zero SD = phenotypes with SD = 0 but AUC $\neq$ 50.0\%. N Wide CI = phenotypes with CI width $>$ 40. N Neg CI = phenotypes where CI lower bound $<$ 0, a mathematical artefact of symmetric interval estimation under small sample sizes. Zero-SD and collapse cases represent prediction failure rather than genuine stability.}
\label{tab:stability}
\resizebox{\textwidth}{!}{%
\begin{tabular}{lrrrrrrrrr}
\hline
\textbf{Tool} & \textbf{N} & \textbf{Mean AUC (\%)} & \textbf{Median AUC (\%)} & \textbf{Mean CI Width} & \textbf{N CI$<$20 (\%)} & \textbf{N Collapse} & \textbf{N Zero SD} & \textbf{N Wide CI} & \textbf{N Neg CI} \\
\hline
Machine Learning & 80 & 66.19 & 64.30 & 30.91 & 20 (25.0\%) & 0  & 0 & 14 & 0 \\
Deep Learning    & 80 & 65.29 & 64.60 & 27.50 & 27 (33.8\%) & 0  & 1 & 13 & 0 \\
Plink            & 79 & 66.63 & 65.71 & 31.37 & 23 (29.1\%) & 0  & 0 & 26 & 0 \\
PRSice           & 79 & 48.01 & 48.57 & 39.95 & 11 (13.9\%) & 0  & 1 & 36 & 5 \\
Lassosum         & 79 & 64.37 & 63.44 & 26.71 & 24 (30.4\%) & 13 & 0 & 18 & 0 \\
\hline
\end{tabular}%
}
\end{table*}

\begin{table*}[!ht]
\centering
\caption{\textbf{Margin-based comparison of workflow families across 80 phenotypes.} The analysis was restricted to phenotypes with both ML and DL results available. Missing values for Plink, PRSice, and Lassosum were set to 0 for consistency. For each phenotype, margins were computed as the absolute difference in AUC between the specified workflow families and grouped into five categories: negligible ($<$1), trivial (1--3), small (3--5), moderate (5--10), and meaningful ($>$10) AUC points.}
\label{tab:margins}
\begin{tabular}{llrr}
\hline
\textbf{Comparison} & \textbf{Margin Bin} & \textbf{Count} & \textbf{Percent} \\
\hline
Best ML/DL vs Best PRS & $<$1 (negligible)  & 8  & 10.0 \\
Best ML/DL vs Best PRS & 1--3 (trivial)     & 11 & 13.8 \\
Best ML/DL vs Best PRS & 3--5 (small)       & 13 & 16.2 \\
Best ML/DL vs Best PRS & 5--10 (moderate)   & 18 & 22.5 \\
Best ML/DL vs Best PRS & $>$10 (meaningful) & 30 & 37.5 \\
ML vs DL               & $<$1 (negligible)  & 13 & 16.2 \\
ML vs DL               & 1--3 (trivial)     & 18 & 22.5 \\
ML vs DL               & 3--5 (small)       & 22 & 27.5 \\
ML vs DL               & 5--10 (moderate)   & 18 & 22.5 \\
ML vs DL               & $>$10 (meaningful) & 9  & 11.2 \\
Plink vs Best ML/DL    & $<$1 (negligible)  & 7  & 8.8 \\
Plink vs Best ML/DL    & 1--3 (trivial)     & 10 & 12.5 \\
Plink vs Best ML/DL    & 3--5 (small)       & 14 & 17.5 \\
Plink vs Best ML/DL    & 5--10 (moderate)   & 20 & 25.0 \\
Plink vs Best ML/DL    & $>$10 (meaningful) & 29 & 36.2 \\
\hline
\end{tabular}
\end{table*}

\begin{table*}[!ht]
\centering
\caption{\textbf{Paired Wilcoxon signed-rank tests comparing workflow families across 80 phenotypes.} Missing PRS values were set to 0 for consistency. The A$>$B / B$>$A / Tied column reports the number of phenotypes where workflow A achieved higher, lower, or equal AUC compared with workflow B. Effect size $r = |z|/\sqrt{n_{\text{nonzero}}}$ is interpreted as negligible ($r<0.10$), small ($0.10\leq r<0.30$), medium ($0.30\leq r<0.50$), or large ($r\geq0.50$).}
\label{tab:wilcoxon}
\resizebox{\textwidth}{!}{%
\begin{tabular}{lrrrrrr}
\hline
\textbf{Comparison} & \textbf{Mean A (\%)} & \textbf{Mean B (\%)} & \textbf{A$>$B / B$>$A / Tied} & \textbf{W} & \textbf{p} & \textbf{r (Effect)} \\
\hline
ML vs PRSice            & 66.19 & 48.38 & 72 / 8 / 0  & 209.0  & $<$0.0001 & 0.76 (large) \\
DL vs PRSice            & 65.29 & 48.38 & 67 / 12 / 1 & 238.0  & $<$0.0001 & 0.74 (large) \\
ML vs Lassosum          & 66.19 & 63.70 & 45 / 35 / 0 & 1462.0 & 0.4486    & 0.08 (negligible) \\
DL vs Lassosum          & 65.29 & 63.70 & 43 / 37 / 0 & 1525.5 & 0.6504    & 0.05 (negligible) \\
ML vs Plink             & 66.19 & 68.15 & 36 / 44 / 0 & 1299.0 & 0.1237    & 0.17 (small) \\
DL vs Plink             & 65.29 & 68.15 & 29 / 51 / 0 & 1151.5 & 0.0246    & 0.25 (small) \\
ML vs DL                & 66.19 & 65.29 & 38 / 40 / 2 & 1396.5 & 0.4732    & 0.08 (negligible) \\
Best ML/DL vs Best PRS  & 68.12 & 71.34 & 27 / 53 / 0 & 1036.0 & 0.0051    & 0.31 (medium) \\
\hline
\end{tabular}%
}
\end{table*}

\begin{table*}[!ht]
\centering
\caption{\textbf{Family-level performance overview across workflow families.} Mean AUC, median AUC, mean CI width, total phenotype-level wins, and decisive wins (\textgreater{}5 AUC points) are summarised for each workflow family. ML and DL summaries are computed across 80 phenotypes. For PRS workflows, the family-level values reported here are computed across 80 phenotypes, including EarlobeFreeorattached; this differs from Table~\ref{tab:stability}, where PRS summaries are restricted to 79 phenotypes after exclusion of EarlobeFreeorattached for stability analysis.}
\label{tab:familylevel}
\resizebox{\textwidth}{!}{%
\begin{tabular}{lrrrrrr}
\hline
\textbf{Tool Family} & \textbf{Phenotypes} & \textbf{Mean AUC (\%)} & \textbf{Median AUC (\%)} & \textbf{Mean CI Width} & \textbf{Total Wins} & \textbf{Decisive Wins ($>$5)} \\
\hline
Machine Learning & 80 & 66.19 & 64.30 & 30.91 & 13 & 2 \\
Deep Learning    & 80 & 65.29 & 64.60 & 27.50 & 14 & 3 \\
Plink            & 80 & 68.15 & 66.67 & 31.37 & 35 & 11 \\
PRSice           & 80 & 48.38 & 49.78 & 39.95 & 0  & 0 \\
Lassosum         & 80 & 63.70 & 63.37 & 26.71 & 18 & 10 \\
\hline
\end{tabular}%
}
\end{table*}

\end{document}